\documentclass[5p, preprint]{elsarticle}

\pdfoutput=1

\usepackage{amsmath}
\usepackage{bm}
\usepackage{graphicx}
\usepackage[unicode]{hyperref}
\usepackage[latin1]{inputenc}
\usepackage{array}
\usepackage{upgreek}
\usepackage{color}
\usepackage{physics}

\begin{document}

\title{Flux compensation for SQUID-detected Magnetic Resonance Force Microscopy}

\author[lei]{M.~de~Wit}
\author[lei]{G.~Welker}
\author[lei]{F.~G.~Hoekstra}
\author[lei]{T.~H.~Oosterkamp\corref{cor1}}
\ead{oosterkamp@physics.leidenuniv.nl}

\cortext[cor1]{Corresponding author}
\address[lei]{Leiden Institute of Physics, Leiden University, PO Box 9504, 2300 RA Leiden, The Netherlands}

\date{\today}

\begin{abstract}
One of the major challenges in performing SQUID-detected Magnetic Resonance Force Microscopy (MRFM) at milliKelvin temperatures is the crosstalk between the radiofrequency (RF) pulses used for the spin manipulation and the SQUID-based detection mechanism. Here we present an approach based on balancing the flux crosstalk using an on-chip feedback coil coupled to the SQUID. This approach does not require any additional components near the location of the sample, and can therefore be applied to any SQUID-based detection scheme to cancel predictable RF interference. We demonstrate the effectiveness of our approach by showing that we can almost completely negate flux crosstalk with an amplitude of up to several $\Phi_0$. This technical achievement allows for complicated magnetic resonance protocols to be performed at temperatures below 50 mK.
\end{abstract}


\maketitle

\section{Introduction}
Magnetic Resonance Force Microscopy (MRFM) is a technique intended for nanoscale Magnetic Resonance Imaging (MRI) \cite{sidles1995,mamin2007,degen2009}. It is based on measuring the forces between spins in the sample and a small magnetic particle attached to the end of a soft cantilever (in the magnet-on-cantilever geometry) \cite{longenecker2012}. A variety of radiofrequency (RF) pulses can be used to manipulate the spins in the sample to generate the signal \cite{rugar2004,garner2004}. The motion of the cantilever, which contains the spin signal, is typically read-out using laser interferometry \cite{poggio2010}.

\begin{figure*}
	\centering
	\includegraphics[width=\textwidth]{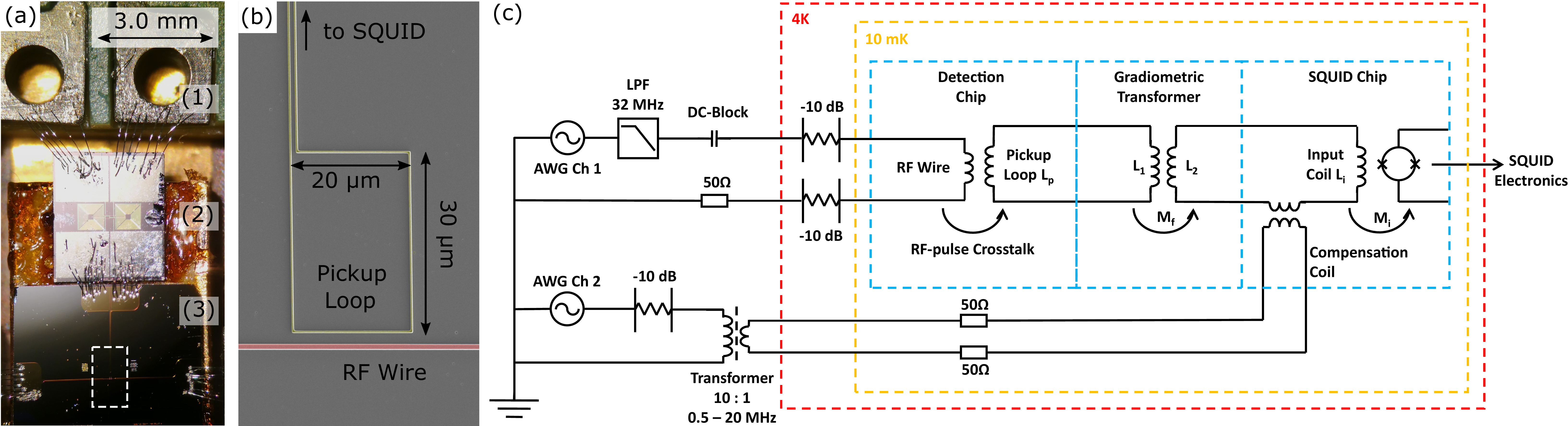}
	\caption{(a) Optical microscope image of (1) the niobium terminals connected to the SQUID input coil, (2) the gradiometric transformer used for the impedance matching, and (3) the MRFM detection chip. The white dashed boxed indicates the location of the zoomed-in image shown in (b) (not to scale). (b) Scanning electron microscope image of the detection chip, showing the NbTiN pickup loop (yellow) and RF wire (red). The pickup loop is connected to the transformer and SQUID as indicated in the schematic shown in (c). (c) Schematic of the electronic circuit used for the flux compensation. The red and yellow dashed boxes indicate different stages of the cryostat, the blue dashed boxes indicate the detection chip, transformer chip, and SQUID chip, all connected using Al-Si(1\%) wirebonds.}
	\label{figure:Setup}
\end{figure*}

The fundamental force sensitivity of MRFM is determined by the thermal force noise. Therefore, one would like to operate the MRFM at the lowest possible temperatures. In order to prevent heating, we use a superconducting microwire as the source for the RF pulses. Furthermore, we have replaced the laser interferometer, which heats the cantilever and sample \cite{poggio2007a}, by a SQUID-based detection scheme \cite{usenko2011}. In this scheme, the motion of the cantilever is determined by measuring the flux induced by the magnetic particle in a superconducting pickup loop, which is coupled to the input coil of a SQUID. A photograph of the experimental setup used in this detection scheme is shown in Fig. \ref{figure:Setup}(a), with a zoom-in on the MRFM detection chip containing the RF wire and pickup loops shown in Fig. \ref{figure:Setup}(b). Due to these adjustments, the SQUID-detected MRFM can be operated at experimentally verified temperatures below 50 mK \cite{vinante2011,wagenaar2016}.

The extreme sensitivity of the SQUID that we rely on to measure the sub-nanometer motion of the cantilever also has a disadvantage: SQUIDs are notoriously sensitive to electromagnetic interference \cite{koch1994,muck1995}. Interference of sufficient intensity reduces the SQUID modulation depth, i.e. the extent to which the SQUID responds to an applied flux. The time-dependent modulation voltage of a SQUID subjected to a low-frequency applied flux $\Phi_a$ and additional RF interference with amplitude $\Phi_{RF}$ is given by \cite{clarke2006}:
\begin{equation}
V(t) = \Delta V_0 \cos\left( \frac{2 \pi \Phi_a(t)}{\Phi_0} \right) ~ J_0 \left( \frac{2\pi \Phi_{RF}}{\Phi_0} \right),
\end{equation}
with $\Delta V_0$ the peak-to-peak modulation depth without RF interference, $J_0$ the zeroth order Bessel function, and $\mathrm{\Phi}$\textsubscript{0} = 2.067 $\cdot 10^{-15}$ Wb, the magnetic flux quantum. The reduced SQUID modulation results in an increase in the measured SQUID noise floor. RF interference that originates from environmental sources can be reduced by using magnetic shielding. However, in the MRFM experiment there is also a local source of RF interference which cannot be avoided by using shielding: the direct crosstalk between the RF pulse and the pickup loop, as discussed in the next section. The presence of this crosstalk is detrimental for MRFM experiments, as it prevents measurements of the spin signal during the pulse, an absolute necessity in many MRFM protocols.

This challenge was also encountered and overcome in the field of SQUID-detected NMR, where the high sensitivity of the SQUIDs offers the possibility to work at very low fields and low frequencies \cite{webb1977,fan1989,greenberg1998,clarke2007}. In order to protect the SQUID from RF interference, a variety of solutions have been developed, but most can be subdivided into two classes. The first class of solutions involves disabling the SQUID by using diodes or Q-spoilers to block high currents \cite{augustine1998,mcdermott2004,matlachov2004,buckenmaier2017}. This type of solutions is relatively easy to implement, but prevents measurements of the NMR signal during the RF pulse. The second class of solutions is based on sending a copy of the RF pulse with the appropriate phase and amplitude to an additional coil in the detection circuit \cite{webb1977,ehnholm1979,pasquarelli1996,pizzella2001}. This balancing coil is often placed around or near the pickup coil which couples the measured NMR signals to the input coil of the SQUID.

In this work, we decribe the measurement scheme used to remove the crosstalk in our SQUID-detected MRFM setup, where we use an on-chip feedback coil in the SQUID input coil circuit to balance nearly all crosstalk before it reaches the SQUID. We start by explaining the compensation method and calibration of the required balancing pulses. We then demonstrate the effectiveness of our approach by showing the reduction in measured crosstalk in the full MRFM setup. The application of this technique to SQUID-dectected MRFM is vital for the operation of MRFM at milliKelvin temperatures.

\section{Circuit and calibration}
As introduced in the previous section, RF pulses are required to manipulate the spins in the sample, which we generate by sending an alternating current through a superconducting RF wire \cite{vinante2011}. In order to coherently modulate the magnetization of the spins, alternating magnetic fields $B_{RF}$ on the order of several mT are necessary \cite{poggio2007b,nichol2012}. To generate a 1 mT field (in the rotating frame of the spins) at a distance of 1 $\upmu$m from the RF wire, a current with peak amplitude $I$ = 10 mA\textsubscript{pk} is required. Given this current and the geometry of the system, the flux crosstalk between the RF pulse and the pickup loop is given by:
\begin{equation} \label{eq:Phix}
	\Phi_{RF} = \eta_{\Phi} \int_{area} B_{RF}(\vec{r}) ~ dA \approx \eta_{\Phi} ~ w I \frac{\mu_0}{2 \pi} \ln \left( \frac{r_2}{r_1} \right),
\end{equation}
where $w$ = 20 $\upmu$m is the width of the pickup loop, $\mu_0$ is the vacuum permeability, and $r_1$ = 2.5 $\upmu$m and $r_2$  = 32.5 $\upmu$m are the distance between the near and far edges of the pickup loop, respectively. We have assumed that all flux enters via the rectangular pickup loop, as we have minimized the parasitic area enclosed by the supply wires.

$\eta_{\Phi}$ is the efficiency of the coupling between the pickup loop and the input coil of the SQUID. To optimize this efficiency, we use a gradiometric transformer to match the inductances of the pickup loop and the SQUID, see Fig. \ref{figure:Setup}(a). For a single transformer circuit, as shown in Fig. \ref{figure:Setup}(c), the efficiency is given by \cite{website:QuantumDesign}:
\begin{equation}
\eta_{\Phi} = \frac{M_f M_i}{\left( L_p + L_{par} + L_1 \right) \left( L_2 + L_i \right) - M_f^2},
\end{equation}
in which the various inductances $L$ and mutual inductances $M$ are defined in Fig. \ref{figure:Setup}(c). $L_{par}$ is the parasitic inductance within the pickup loop circuit, which is dominated by the wirebonds between the detection chip and the transformer. We use a sub-optimally designed gradiometric transformer to match the impedance of the pickup loop and input coil, resulting in $\eta_{\Phi} \approx$ 3.5 \%. Using this efficiency together with the other experimental parameters, we find from Eq. \ref{eq:Phix} that a 10 mA current induces a crosstalk in the SQUID of $\Phi_{RF} = 3.6 \cdot 10^{-15}$ Wb $\sim$ 1.8 $\mathrm{\Phi}$\textsubscript{0}. Given that the SQUID noise floor at temperatures below 4K is generally less than 1-2 $\upmu \Phi_0 / \sqrt{\mathrm{Hz}}$, this crosstalk is quite significant.

To solve this issue, we use an additional feedback transformer to precisely cancel this crosstalk flux in the SQUID, using the electrical circuit shown in Fig. \ref{figure:Setup}(c). We use a SQUID with an on-chip additional feedback transformer\footnote{Magnicon integrated 2-stage current sensor C70M116W}. A dual-channel arbitrary waveform generator (AWG) is used to send both the RF pulse and the compensation pulse. The first channel is used to send the current to the RF wire. This current is intended to generate the magnetic fields to perform NMR protocols, but also creates unwanted crosstalk flux in the pickup loop. Low temperature attenuators are used to reduce the noise originating from the room temperature electronics and filters. The second channel is used to send a compensation pulse with precisely the correct gain and phase shift to the compensation coil in order to balance the crosstalk of the RF pulse. A ferrite core transformer is used to decouple the highly sensitive feedback transformer from low-frequency noise on the electrical ground of the cryostat. As the compensation coil is so strongly coupled to the input coil of the SQUID, 50 Ohm resistors are used to attenuate the current at the 10 mK plate of the dilution refrigerator to suppress noise currents in the compensation circuit.

\begin{figure}
	\centering
	\includegraphics[width=\columnwidth]{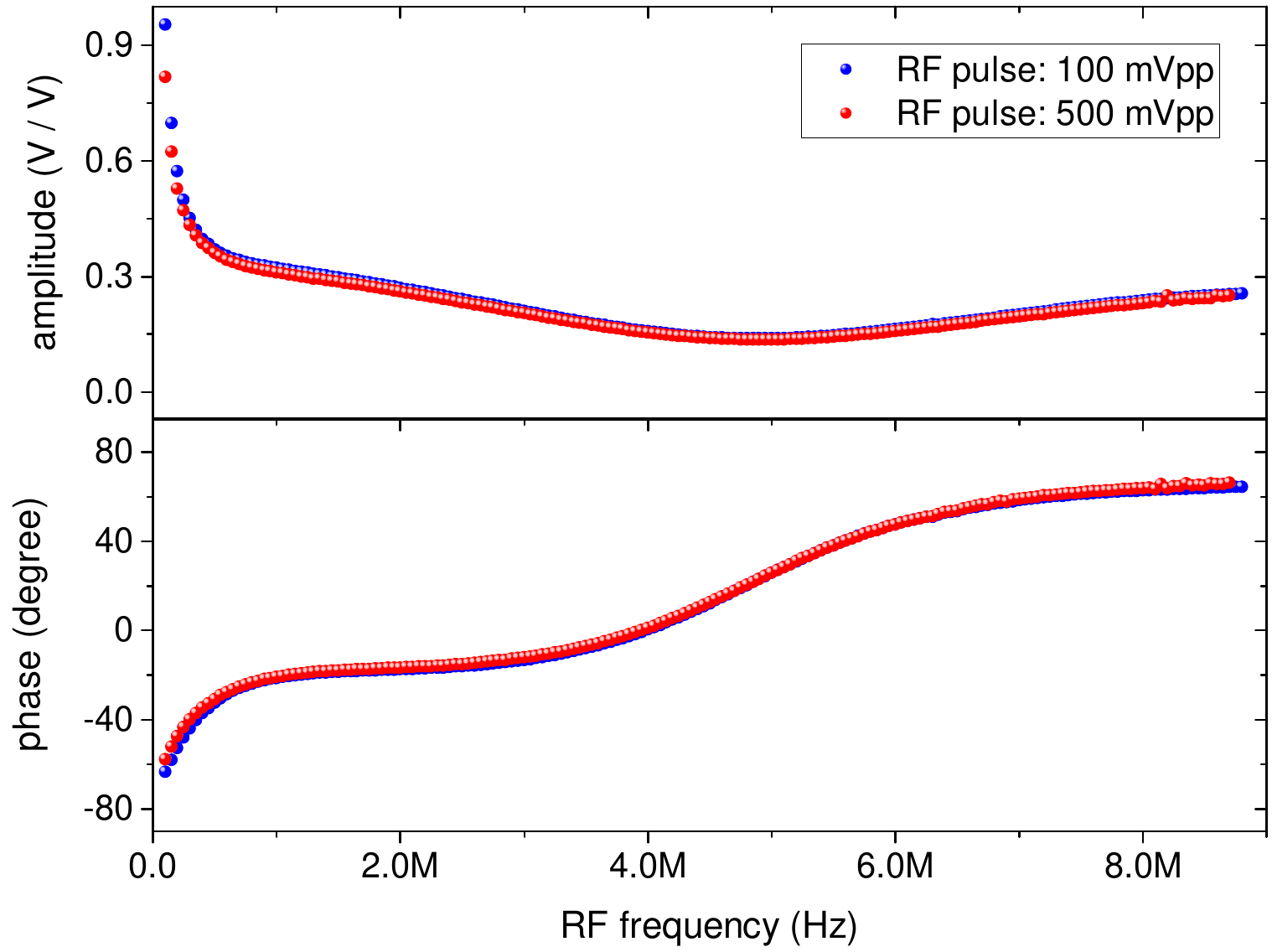}
	\caption{Calibrated amplitude $r(f)$ (top) and phase $\phi(f)$ (bottom) of the compensation pulse for an RF pulse with reference phase 0. Both the amplitude and phase result from the combination of the transfer functions of the RF circuit and the compensation circuit, as shown in Fig. \ref{figure:Setup}(c).}
	\label{figure:Calibration}
\end{figure}

The required gain and phase shift are calibrated by using a lock-in amplifier to measure the crosstalk in the SQUID during a continuous, constant frequency RF signal. The amplitude $r(f)$ and phase $\phi(f)$ of the compensation pulse is varied until a minimum in the measured crosstalk is obtained. Because of the frequency dependence of the transfer functions of the various circuits, this calibration must be repeated for the full RF pulse frequency range required for the experiments, the result of which is shown in Fig. \ref{figure:Calibration}. The blue and red curves are measured for different RF signal amplitudes. The good correspondence between the two shows that the SQUID does not suffer from large non-linearities in this range.

\begin{figure}
	\centering
	\includegraphics[width=\columnwidth]{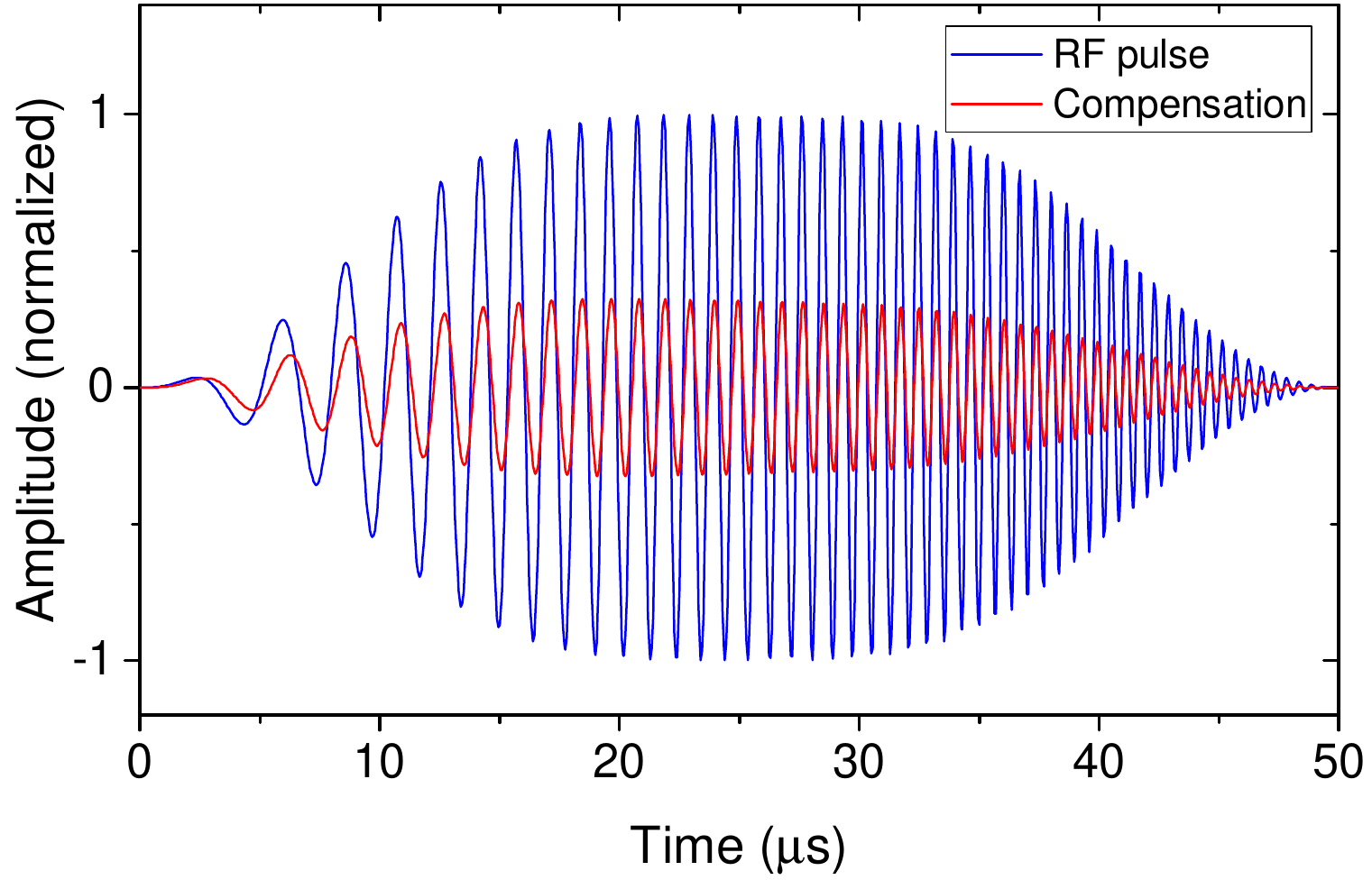}
	\caption{Example of a complex RF pulse, typically used for ARP protocols in MRFM, plus pre-calculated compensation pulse using the Fourier transform method. The RF pulse consists of a WURST amplitude envelope and a linear frequency chirp from 100 kHz to 1.9 MHz.}
	\label{figure:ARP}
\end{figure}

It is straightforward to use the calibration from Fig. \ref{figure:Calibration} to properly compensate the flux from pulses consisting of a single frequency, as required for e.g. saturation experiments \cite{wagenaar2016}. However, it can also be used to compensate for the crosstalk from more complex RF pulses, such as the pulses required for adiabatic rapid passage (ARP), the technique used to coherently flip the magnetization of a spin ensemble \cite{poggio2007b,garwood2001}. An ARP pulse consists of a frequency-chirp combined with an amplitude envelope, an example of which is given by the blue curve in Fig. \ref{figure:ARP}. In this particular example, the amplitude envelope is of the WURST kind \cite{odell2013,claridge2016}, given by $A(t) = 1 - \abs{ \cos{\left( \frac{\pi t}{t_p} \right) }}^4$, where the pulse starts at $t$ = 0 and ends at $t$ = $t_p$.

In order to find the appropriate compensation pulse for an arbitrary RF pulse, we take the discrete Fourier transform of the RF pulse, and multiply each frequency component with the corresponding calibrated amplitude and phase represented as the complex number $z(f) = r(f) e^{i\phi(f)}$. The required compensation pulse is obtained by taking the inverse Fourier transform to return to the time-domain. The resulting compensation pulse for the example ARP pulse is shown in red in Fig. \ref{figure:ARP}.

\section{Results}
\begin{figure}
	\centering
	\includegraphics[width=\columnwidth]{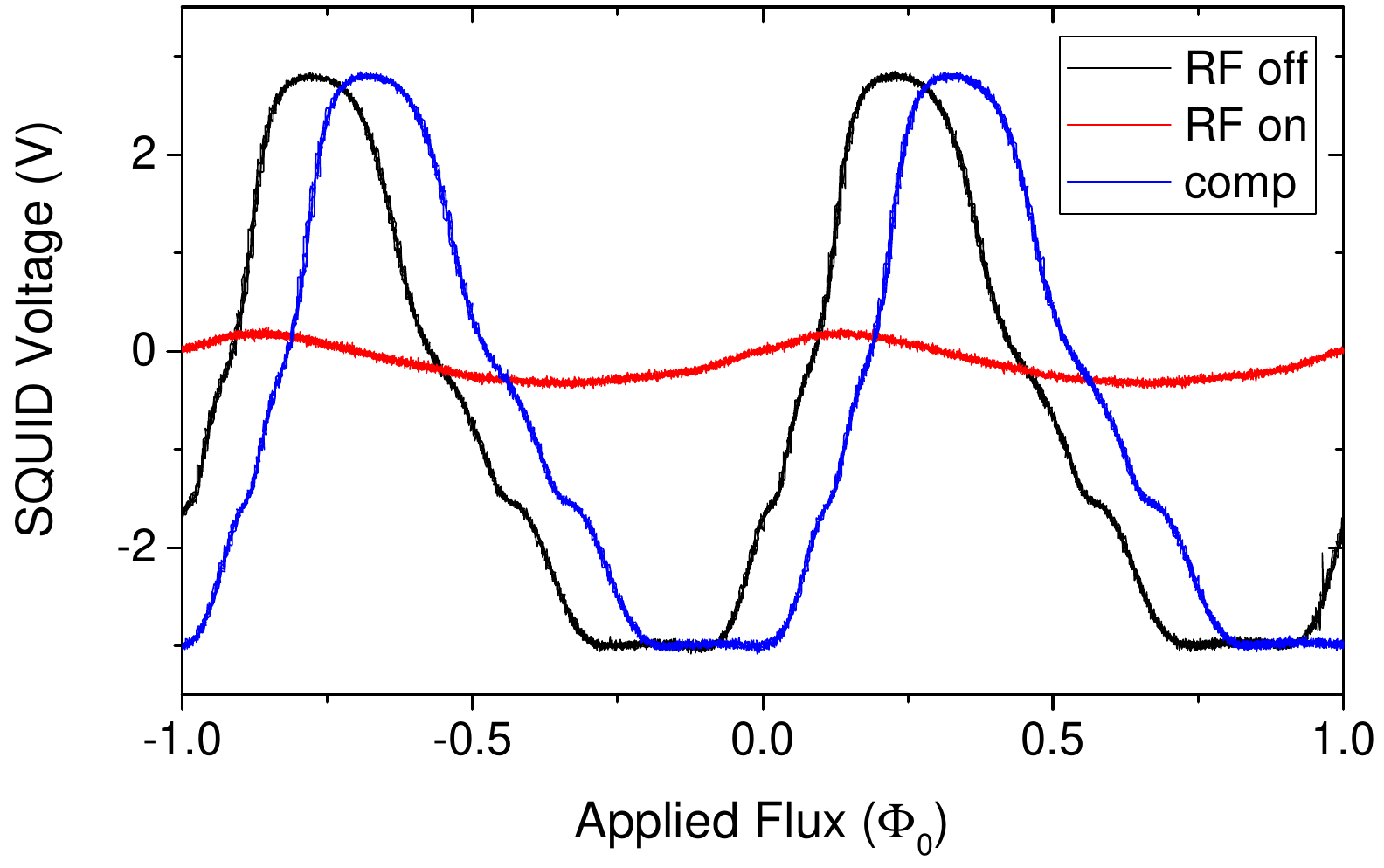}
	\caption{Effects of a 1.908 MHz, 0.88 $\Phi_0$ pulse on the SQUID modulation. The modulation without RF field is shown in black. The red curve shows the suppressed modulation with an unbalanced RF flux crosstalk, while the blue curve shows the restored modulation with optimized compensation.}
	\label{figure:Mod}
\end{figure}

To demonstrate the effectiveness of the flux crosstalk compensation method, we show the SQUID modulation in Fig. \ref{figure:Mod}. We apply a test flux $\Phi_a \sim$ 2 $\mathrm{\Phi}$\textsubscript{0} in the SQUID at a frequency of 23 Hz. The reference modulation without RF pulse is shown in black, where we find a SQUID modulation depth of 5.9 V\textsubscript{pp}. When switching on the RF pulse with a frequency of 1.908 MHz and a peak-to-peak amplitude of 0.88 $\mathrm{\Phi}$\textsubscript{0}, the modulation depth is significantly reduced, as can be seen from the red curve. The SQUID's noise susceptibility, which can be quantified by looking at the maximum slope of the modulation, is increased by a factor of 8. By sending the suitable compensation pulse, we are able to restore the SQUID modulation, as shown by the blue curve. The compensation, and corresponding canceling of the flux crosstalk in the SQUID input circuit, leads to a recovery of the SQUID noise level to within 3\% of that without RF pulse. Note that the different modulations are manually shifted by a random phase. This does not influence the actual experiment.

\begin{figure}
	\centering
	\includegraphics[width=\columnwidth]{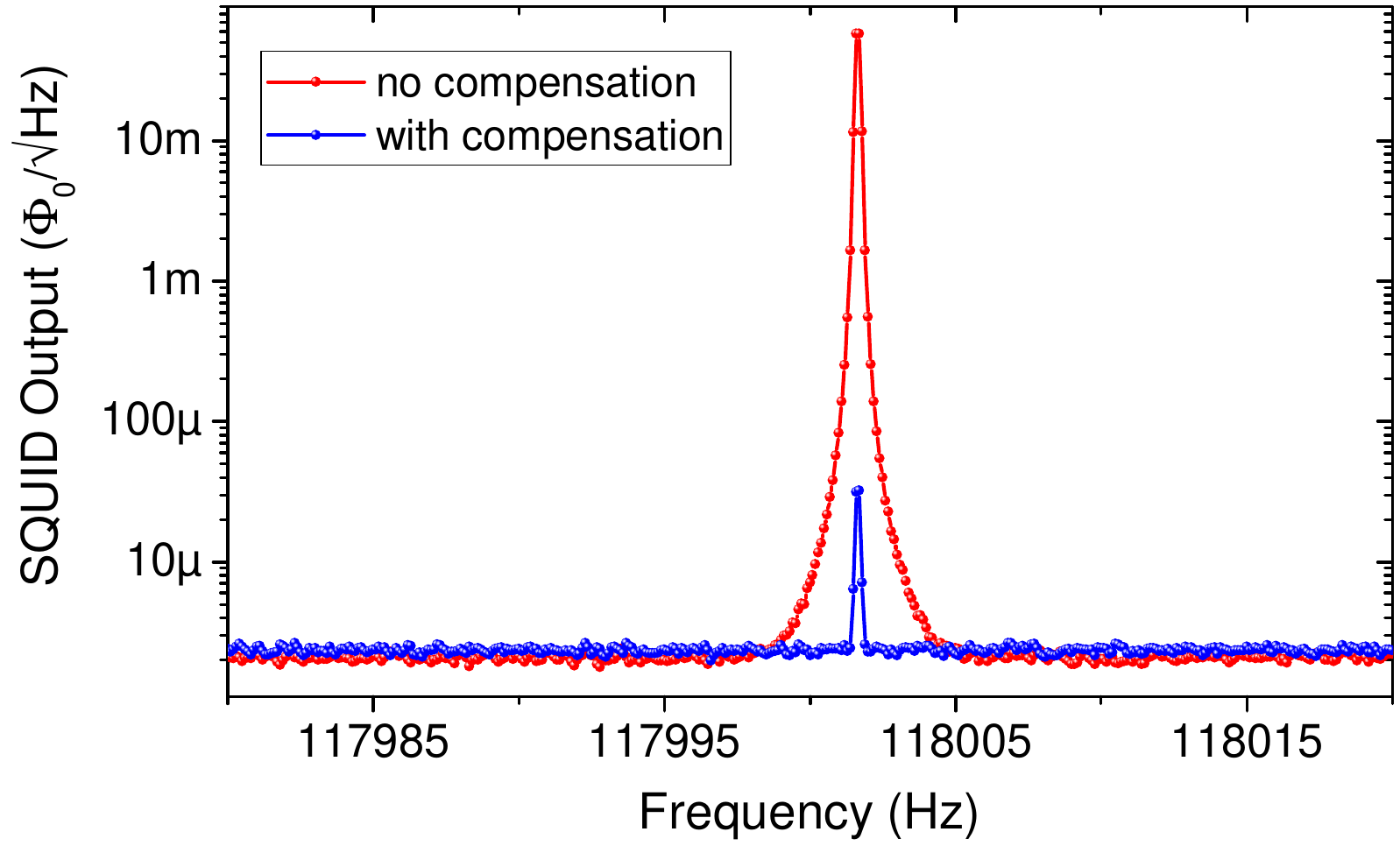}
	\caption{Demonstration of the reduction of crosstalk by compensating the RF pulse. The measured flux crosstalk has been reduced from 26 m$\mathrm{\Phi}$\textsubscript{0} without compensation to less than 15 $\upmu \mathrm{\Phi}$\textsubscript{0} with compensation.}
	\label{figure:Crosstalk}
\end{figure}

The previous experiment of the SQUID modulation depth gives an idea of the effect of the RF pulse on the SQUID noise floor, i.e. the ability to measure at frequencies not equal to the frequency of the RF pulse. A direct visualization of the effect of the compensation at the RF pulse frequency is shown in Fig. \ref{figure:Crosstalk}, where we see a small part of the SQUID spectra during the application of a 118 kHz, 0.3 mA\textsubscript{pp} RF pulse, with and without compensation. Each spectrum has been averaged 100 times with a total measurement time of 1000 seconds. The measured integrated flux crosstalk has been reduced from 74 m$\mathrm{\Phi}$\textsubscript{0,pp} without compensation to below 42 $\upmu \mathrm{\Phi}$\textsubscript{0,pp} with compensation. Thus, the crosstalk has been reduced to less than 0.1\% of the uncompensated value. The remaining flux crosstalk is the result of a small drift in the transfer functions of the RF wire or compensation circuits due to heating of the low temperature attenuators. We expect that this problem is reduced for pulses of shorter duration. Note that in order to reach these levels of crosstalk reduction, the compensation pulse amplitude has to be calibrated to an accuracy better than 0.1\%, and the phase to better than 0.1 degree.

The RF frequency of 118 kHz was selected in order to prevent aliasing in the data acquisition. However, note that the compensation scheme can in principle be applied over a large bandwidth, from DC up to at least tens of MHz. In the current experiment, this bandwidth is limited by the DC block in the RF circuit, and the bandwidth of the SQUID feedback of about 20 MHz.

Note that the compensation scheme is suitable for low temperature operation, as it generates only very little dissipation. Compensating a 10 mA\textsubscript{pk} RF current requires a balancing current of about 200 $\upmu$A\textsubscript{pk} in the low temperature calibration coil circuit. Given that this current dissipates over the two 50 Ohm resistors, this leads to a power dissipation of less than 2 $\upmu$W at the 10 mK plate. This power is sufficiently small that this will not significantly heat up the 10 mK plate.

\section{Conclusions and outlook}
We have presented a crosstalk compensation scheme that is easy to implement without any local adjustments near the sample or pickup loops. The compensation scheme allows for relatively strong RF pulses without any adverse effect on the SQUID sensitivity. This means that data acquisition with the SQUID does not have to be interrupted or compromised during RF pulses. This is a vital requirement for MRFM experiments, and initially was considered one of the major arguments against a SQUID-detected MRFM setup. Furthermore, even though our approach is very similar to what is done in the SQUID-detected NMR community, our approach to balance the crosstalk at the location of the SQUID instead of near the sample could allow for a broader application of the balancing technique.

An extended application would be to use this scheme to cancel noise in an applied external magnetic field. An external magnetic field in MRFM is useful due to the enhancement of the Boltzmann polarization, leading to a larger signal for the same volume of spins. However, due to the extreme sensitivity to magnetic field noise of the SQUID, applying an external magnetic field with sufficient stability is not an easy task. Assuming we have a gradiometric SQUID with a parasitic area of only 1 $\upmu$m$^2$, applying a 1 T external magnetic field without degradation of the SQUID noise requires a field stability of 1 part per billion. This is far beyond the resolution of present-day magnet power supply systems, which is of the order of 10 ppm \cite{choi2016}. Placing the SQUID outside of the magnetic field, and instead using a gradiometric pickup coil with a coupling efficiency of 1 \% reduces the stability demand to 0.1 ppm, which is still beyond reach. A potential solution would be to use a persistent current switch, but achieving this at 10 mK is still technologically challenging \cite{waarde2016}. Alternatively, one could redirect a part of the current from the external magnetic field to the compensation coil, after proper attenuation and phase shifting. Any noise in the power supply of the external magnet will now be compensated at the SQUID level. When the current from the magnet power supply is low-pass filtered to a bandwidth of 10 Hz, even a delay of 100 $\upmu$s is acceptable to obtain a noise reduction of over a factor of 100. In combination with the other proposed solutions this should be sufficient to operate the SQUID in an external field without a reduction in sensitivity.

The possibility to continue to measure with the SQUID even during an RF protocol opens the way to perform MRFM experiments that rely on continuous application of ARP pulses at low temperatures. The fundamental limit of sensitivity of an MRFM experiment is dictated by the thermal force noise $\sqrt{S_{th}} = \sqrt{4 k_B T \gamma ~ BW}$, with $\gamma$ the damping of the cantilever and $BW$ the measurement bandwidth. Thus, lower operating temperatures in principle allow for measurements on smaller spin ensembles, and would be a new step towards the imaging of a single nuclear spin.

\section{acknowledgement}
The authors thank M. Camp, K. Heeck, G. Koning, J.P. Koning, and L. Crama for technical support. The authors thank D.J. Thoen, T.M. Klapwijk, and A. Endo for providing us with the NbTiN and assistance in the fabrication of the detection chip. The authors thank T.H.A. van der Reep for proofreading the manuscript. This work is supported by the Netherlands Organisation for Scientific Research (NWO) through a VICI fellowship to T.H.O., and through the Nanofront program.

\bibliography{SQUID_Comp_Literature}

\begin{thebibliography}{33}%
\makeatletter
\providecommand \@ifxundefined [1]{%
 \@ifx{#1\undefined}
}%
\providecommand \@ifnum [1]{%
 \ifnum #1\expandafter \@firstoftwo
 \else \expandafter \@secondoftwo
 \fi
}%
\providecommand \@ifx [1]{%
 \ifx #1\expandafter \@firstoftwo
 \else \expandafter \@secondoftwo
 \fi
}%
\providecommand \natexlab [1]{#1}%
\providecommand \enquote  [1]{``#1''}%
\providecommand \bibnamefont  [1]{#1}%
\providecommand \bibfnamefont [1]{#1}%
\providecommand \citenamefont [1]{#1}%
\providecommand \href@noop [0]{\@secondoftwo}%
\providecommand \href [0]{\begingroup \@sanitize@url \@href}%
\providecommand \@href[1]{\@@startlink{#1}\@@href}%
\providecommand \@@href[1]{\endgroup#1\@@endlink}%
\providecommand \@sanitize@url [0]{\catcode `\\12\catcode `\$12\catcode
  `\&12\catcode `\#12\catcode `\^12\catcode `\_12\catcode `\%12\relax}%
\providecommand \@@startlink[1]{}%
\providecommand \@@endlink[0]{}%
\providecommand \url  [0]{\begingroup\@sanitize@url \@url }%
\providecommand \@url [1]{\endgroup\@href {#1}{\urlprefix }}%
\providecommand \urlprefix  [0]{URL }%
\providecommand \Eprint [0]{\href }%
\providecommand \doibase [0]{http://dx.doi.org/}%
\providecommand \selectlanguage [0]{\@gobble}%
\providecommand \bibinfo  [0]{\@secondoftwo}%
\providecommand \bibfield  [0]{\@secondoftwo}%
\providecommand \translation [1]{[#1]}%
\providecommand \BibitemOpen [0]{}%
\providecommand \bibitemStop [0]{}%
\providecommand \bibitemNoStop [0]{.\EOS\space}%
\providecommand \EOS [0]{\spacefactor3000\relax}%
\providecommand \BibitemShut  [1]{\csname bibitem#1\endcsname}%
\let\auto@bib@innerbib\@empty
\bibitem [{\citenamefont {Sidles}\ \emph {et~al.}(1995)\citenamefont {Sidles},
  \citenamefont {Garbini}, \citenamefont {Bruland}, \citenamefont {Rugar},
  \citenamefont {Z{\"u}ger}, \citenamefont {Hoen},\ and\ \citenamefont
  {Yannoni}}]{sidles1995}%
  \BibitemOpen
  \bibfield  {author} {\bibinfo {author} {\bibfnamefont {J.~A.}\ \bibnamefont
  {Sidles}}, \bibinfo {author} {\bibfnamefont {J.~L.}\ \bibnamefont {Garbini}},
  \bibinfo {author} {\bibfnamefont {K.~J.}\ \bibnamefont {Bruland}}, \bibinfo
  {author} {\bibfnamefont {D.}~\bibnamefont {Rugar}}, \bibinfo {author}
  {\bibfnamefont {O.}~\bibnamefont {Z{\"u}ger}}, \bibinfo {author}
  {\bibfnamefont {S.}~\bibnamefont {Hoen}}, \ and\ \bibinfo {author}
  {\bibfnamefont {C.~S.}\ \bibnamefont {Yannoni}},\ }\href {\doibase
  10.1103/RevModPhys.67.249} {\bibfield  {journal} {\bibinfo  {journal}
  {Reviews of Modern Physics}\ }\textbf {\bibinfo {volume} {67}},\ \bibinfo
  {pages} {249} (\bibinfo {year} {1995})}\BibitemShut {NoStop}%
\bibitem [{\citenamefont {Mamin}\ \emph {et~al.}(2007)\citenamefont {Mamin},
  \citenamefont {Poggio}, \citenamefont {Degen},\ and\ \citenamefont
  {Rugar}}]{mamin2007}%
  \BibitemOpen
  \bibfield  {author} {\bibinfo {author} {\bibfnamefont {H.~J.}\ \bibnamefont
  {Mamin}}, \bibinfo {author} {\bibfnamefont {M.}~\bibnamefont {Poggio}},
  \bibinfo {author} {\bibfnamefont {C.~L.}\ \bibnamefont {Degen}}, \ and\
  \bibinfo {author} {\bibfnamefont {D.}~\bibnamefont {Rugar}},\ }\href
  {\doibase 10.1038/nnano.2007.105} {\bibfield  {journal} {\bibinfo  {journal}
  {Nature Nanotechnology}\ }\textbf {\bibinfo {volume} {2}},\ \bibinfo {pages}
  {301} (\bibinfo {year} {2007})}\BibitemShut {NoStop}%
\bibitem [{\citenamefont {Degen}\ \emph {et~al.}(2009)\citenamefont {Degen},
  \citenamefont {Poggio}, \citenamefont {Mamin}, \citenamefont {Rettner},\ and\
  \citenamefont {Rugar}}]{degen2009}%
  \BibitemOpen
  \bibfield  {author} {\bibinfo {author} {\bibfnamefont {C.~L.}\ \bibnamefont
  {Degen}}, \bibinfo {author} {\bibfnamefont {M.}~\bibnamefont {Poggio}},
  \bibinfo {author} {\bibfnamefont {H.~J.}\ \bibnamefont {Mamin}}, \bibinfo
  {author} {\bibfnamefont {C.~T.}\ \bibnamefont {Rettner}}, \ and\ \bibinfo
  {author} {\bibfnamefont {D.}~\bibnamefont {Rugar}},\ }\href {\doibase
  10.1073/pnas.0812068106} {\bibfield  {journal} {\bibinfo  {journal}
  {Proceedings of the National Academy of Sciences}\ }\textbf {\bibinfo
  {volume} {106}},\ \bibinfo {pages} {1313} (\bibinfo {year}
  {2009})}\BibitemShut {NoStop}%
\bibitem [{\citenamefont {Longenecker}\ \emph {et~al.}(2012)\citenamefont
  {Longenecker}, \citenamefont {Mamin}, \citenamefont {Senko}, \citenamefont
  {Chen}, \citenamefont {Rettner}, \citenamefont {Rugar},\ and\ \citenamefont
  {Marohn}}]{longenecker2012}%
  \BibitemOpen
  \bibfield  {author} {\bibinfo {author} {\bibfnamefont {J.~G.}\ \bibnamefont
  {Longenecker}}, \bibinfo {author} {\bibfnamefont {H.~J.}\ \bibnamefont
  {Mamin}}, \bibinfo {author} {\bibfnamefont {A.~W.}\ \bibnamefont {Senko}},
  \bibinfo {author} {\bibfnamefont {L.}~\bibnamefont {Chen}}, \bibinfo {author}
  {\bibfnamefont {C.~T.}\ \bibnamefont {Rettner}}, \bibinfo {author}
  {\bibfnamefont {D.}~\bibnamefont {Rugar}}, \ and\ \bibinfo {author}
  {\bibfnamefont {J.~A.}\ \bibnamefont {Marohn}},\ }\href {\doibase
  10.1021/nn3030628} {\bibfield  {journal} {\bibinfo  {journal} {ACS Nano}\
  }\textbf {\bibinfo {volume} {6}},\ \bibinfo {pages} {9637} (\bibinfo {year}
  {2012})}\BibitemShut {NoStop}%
\bibitem [{\citenamefont {Rugar}\ \emph {et~al.}(2004)\citenamefont {Rugar},
  \citenamefont {Budakian}, \citenamefont {Mamin},\ and\ \citenamefont
  {Chui}}]{rugar2004}%
  \BibitemOpen
  \bibfield  {author} {\bibinfo {author} {\bibfnamefont {D.}~\bibnamefont
  {Rugar}}, \bibinfo {author} {\bibfnamefont {R.}~\bibnamefont {Budakian}},
  \bibinfo {author} {\bibfnamefont {H.~J.}\ \bibnamefont {Mamin}}, \ and\
  \bibinfo {author} {\bibfnamefont {B.~W.}\ \bibnamefont {Chui}},\ }\href
  {\doibase 10.1038/nature02658} {\bibfield  {journal} {\bibinfo  {journal}
  {Nature}\ }\textbf {\bibinfo {volume} {430}},\ \bibinfo {pages} {329}
  (\bibinfo {year} {2004})}\BibitemShut {NoStop}%
\bibitem [{\citenamefont {Garner}\ \emph {et~al.}(2004)\citenamefont {Garner},
  \citenamefont {Kuehn}, \citenamefont {Dawlaty}, \citenamefont {Jenkins},\
  and\ \citenamefont {Marohn}}]{garner2004}%
  \BibitemOpen
  \bibfield  {author} {\bibinfo {author} {\bibfnamefont {S.~R.}\ \bibnamefont
  {Garner}}, \bibinfo {author} {\bibfnamefont {S.}~\bibnamefont {Kuehn}},
  \bibinfo {author} {\bibfnamefont {J.~M.}\ \bibnamefont {Dawlaty}}, \bibinfo
  {author} {\bibfnamefont {N.~E.}\ \bibnamefont {Jenkins}}, \ and\ \bibinfo
  {author} {\bibfnamefont {J.~A.}\ \bibnamefont {Marohn}},\ }\href {\doibase
  10.1063/1.1762700} {\bibfield  {journal} {\bibinfo  {journal} {Applied
  Physics Letters}\ }\textbf {\bibinfo {volume} {84}},\ \bibinfo {pages} {5091}
  (\bibinfo {year} {2004})}\BibitemShut {NoStop}%
\bibitem [{\citenamefont {Poggio}\ and\ \citenamefont
  {Degen}(2010)}]{poggio2010}%
  \BibitemOpen
  \bibfield  {author} {\bibinfo {author} {\bibfnamefont {M.}~\bibnamefont
  {Poggio}}\ and\ \bibinfo {author} {\bibfnamefont {C.~L.}\ \bibnamefont
  {Degen}},\ }\href {\doibase 10.1088/0957-4484/21/34/342001} {\bibfield
  {journal} {\bibinfo  {journal} {Nanotechnology}\ }\textbf {\bibinfo {volume}
  {21}},\ \bibinfo {pages} {342001} (\bibinfo {year} {2010})}\BibitemShut
  {NoStop}%
\bibitem [{\citenamefont {Poggio}\ \emph
  {et~al.}(2007{\natexlab{a}})\citenamefont {Poggio}, \citenamefont {Degen},
  \citenamefont {Mamin},\ and\ \citenamefont {Rugar}}]{poggio2007a}%
  \BibitemOpen
  \bibfield  {author} {\bibinfo {author} {\bibfnamefont {M.}~\bibnamefont
  {Poggio}}, \bibinfo {author} {\bibfnamefont {C.~L.}\ \bibnamefont {Degen}},
  \bibinfo {author} {\bibfnamefont {H.~J.}\ \bibnamefont {Mamin}}, \ and\
  \bibinfo {author} {\bibfnamefont {D.}~\bibnamefont {Rugar}},\ }\href
  {\doibase 10.1103/PhysRevLett.99.017201} {\bibfield  {journal} {\bibinfo
  {journal} {Physical Review Letters}\ }\textbf {\bibinfo {volume} {99}},\
  \bibinfo {pages} {017201} (\bibinfo {year} {2007}{\natexlab{a}})}\BibitemShut
  {NoStop}%
\bibitem [{\citenamefont {Usenko}\ \emph {et~al.}(2011)\citenamefont {Usenko},
  \citenamefont {Vinante}, \citenamefont {Wijts},\ and\ \citenamefont
  {Oosterkamp}}]{usenko2011}%
  \BibitemOpen
  \bibfield  {author} {\bibinfo {author} {\bibfnamefont {O.}~\bibnamefont
  {Usenko}}, \bibinfo {author} {\bibfnamefont {A.}~\bibnamefont {Vinante}},
  \bibinfo {author} {\bibfnamefont {G.~H. C.~J.}\ \bibnamefont {Wijts}}, \ and\
  \bibinfo {author} {\bibfnamefont {T.~H.}\ \bibnamefont {Oosterkamp}},\ }\href
  {\doibase 10.1063/1.3570628} {\bibfield  {journal} {\bibinfo  {journal}
  {Applied Physics Letters}\ }\textbf {\bibinfo {volume} {98}},\ \bibinfo
  {pages} {133105} (\bibinfo {year} {2011})}\BibitemShut {NoStop}%
\bibitem [{\citenamefont {Vinante}\ \emph {et~al.}(2011)\citenamefont
  {Vinante}, \citenamefont {Wijts}, \citenamefont {Usenko}, \citenamefont
  {Schinkelshoek},\ and\ \citenamefont {Oosterkamp}}]{vinante2011}%
  \BibitemOpen
  \bibfield  {author} {\bibinfo {author} {\bibfnamefont {A.}~\bibnamefont
  {Vinante}}, \bibinfo {author} {\bibfnamefont {G.}~\bibnamefont {Wijts}},
  \bibinfo {author} {\bibfnamefont {O.}~\bibnamefont {Usenko}}, \bibinfo
  {author} {\bibfnamefont {L.}~\bibnamefont {Schinkelshoek}}, \ and\ \bibinfo
  {author} {\bibfnamefont {T.~H.}\ \bibnamefont {Oosterkamp}},\ }\href
  {\doibase 10.1038/ncomms1581} {\bibfield  {journal} {\bibinfo  {journal}
  {Nature Communications}\ }\textbf {\bibinfo {volume} {2}},\ \bibinfo {pages}
  {572} (\bibinfo {year} {2011})}\BibitemShut {NoStop}%
\bibitem [{\citenamefont {Wagenaar}\ \emph {et~al.}(2016)\citenamefont
  {Wagenaar}, \citenamefont {den Haan}, \citenamefont {de~Voogd}, \citenamefont
  {Bossoni}, \citenamefont {de~Jong}, \citenamefont {de~Wit}, \citenamefont
  {Bastiaans}, \citenamefont {Thoen}, \citenamefont {Endo}, \citenamefont
  {Klapwijk}, \citenamefont {Zaanen},\ and\ \citenamefont
  {Oosterkamp}}]{wagenaar2016}%
  \BibitemOpen
  \bibfield  {author} {\bibinfo {author} {\bibfnamefont {J.~J.~T.}\
  \bibnamefont {Wagenaar}}, \bibinfo {author} {\bibfnamefont {A.~M.~J.}\
  \bibnamefont {den Haan}}, \bibinfo {author} {\bibfnamefont {J.~M.}\
  \bibnamefont {de~Voogd}}, \bibinfo {author} {\bibfnamefont {L.}~\bibnamefont
  {Bossoni}}, \bibinfo {author} {\bibfnamefont {T.~A.}\ \bibnamefont
  {de~Jong}}, \bibinfo {author} {\bibfnamefont {M.}~\bibnamefont {de~Wit}},
  \bibinfo {author} {\bibfnamefont {K.~M.}\ \bibnamefont {Bastiaans}}, \bibinfo
  {author} {\bibfnamefont {D.~J.}\ \bibnamefont {Thoen}}, \bibinfo {author}
  {\bibfnamefont {A.}~\bibnamefont {Endo}}, \bibinfo {author} {\bibfnamefont
  {T.~M.}\ \bibnamefont {Klapwijk}}, \bibinfo {author} {\bibfnamefont
  {J.}~\bibnamefont {Zaanen}}, \ and\ \bibinfo {author} {\bibfnamefont {T.~H.}\
  \bibnamefont {Oosterkamp}},\ }\href {\doibase
  10.1103/PhysRevApplied.6.014007} {\bibfield  {journal} {\bibinfo  {journal}
  {Physical Review Applied}\ }\textbf {\bibinfo {volume} {6}},\ \bibinfo
  {pages} {014007} (\bibinfo {year} {2016})}\BibitemShut {NoStop}%
\bibitem [{\citenamefont {Koch}\ \emph {et~al.}(1994)\citenamefont {Koch},
  \citenamefont {Foglietti}, \citenamefont {Rozen}, \citenamefont {Stawiasz},
  \citenamefont {Ketchen}, \citenamefont {Lathrop}, \citenamefont {Sun},\ and\
  \citenamefont {Gallagher}}]{koch1994}%
  \BibitemOpen
  \bibfield  {author} {\bibinfo {author} {\bibfnamefont {R.~H.}\ \bibnamefont
  {Koch}}, \bibinfo {author} {\bibfnamefont {V.}~\bibnamefont {Foglietti}},
  \bibinfo {author} {\bibfnamefont {J.~R.}\ \bibnamefont {Rozen}}, \bibinfo
  {author} {\bibfnamefont {K.~G.}\ \bibnamefont {Stawiasz}}, \bibinfo {author}
  {\bibfnamefont {M.~B.}\ \bibnamefont {Ketchen}}, \bibinfo {author}
  {\bibfnamefont {D.~K.}\ \bibnamefont {Lathrop}}, \bibinfo {author}
  {\bibfnamefont {J.~Z.}\ \bibnamefont {Sun}}, \ and\ \bibinfo {author}
  {\bibfnamefont {W.~J.}\ \bibnamefont {Gallagher}},\ }\href {\doibase
  10.1063/1.113046} {\bibfield  {journal} {\bibinfo  {journal} {Applied Physics
  Letters}\ }\textbf {\bibinfo {volume} {65}},\ \bibinfo {pages} {100}
  (\bibinfo {year} {1994})}\BibitemShut {NoStop}%
\bibitem [{\citenamefont {M{\"u}ck}\ \emph {et~al.}(1995)\citenamefont
  {M{\"u}ck}, \citenamefont {Dechert}, \citenamefont {Gail}, \citenamefont
  {Kreutzbruck}, \citenamefont {Sch{\"o}ne},\ and\ \citenamefont
  {Weidl}}]{muck1995}%
  \BibitemOpen
  \bibfield  {author} {\bibinfo {author} {\bibfnamefont {M.}~\bibnamefont
  {M{\"u}ck}}, \bibinfo {author} {\bibfnamefont {J.}~\bibnamefont {Dechert}},
  \bibinfo {author} {\bibfnamefont {J.}~\bibnamefont {Gail}}, \bibinfo {author}
  {\bibfnamefont {M.}~\bibnamefont {Kreutzbruck}}, \bibinfo {author}
  {\bibfnamefont {S.}~\bibnamefont {Sch{\"o}ne}}, \ and\ \bibinfo {author}
  {\bibfnamefont {R.}~\bibnamefont {Weidl}},\ }\href {\doibase
  10.1063/1.1145308} {\bibfield  {journal} {\bibinfo  {journal} {Review of
  Scientific Instruments}\ }\textbf {\bibinfo {volume} {66}},\ \bibinfo {pages}
  {4690} (\bibinfo {year} {1995})}\BibitemShut {NoStop}%
\bibitem [{\citenamefont {Clarke}\ and\ \citenamefont
  {Braginski}(2006)}]{clarke2006}%
  \BibitemOpen
  \bibfield  {author} {\bibinfo {author} {\bibfnamefont {J.}~\bibnamefont
  {Clarke}}\ and\ \bibinfo {author} {\bibfnamefont {A.~I.}\ \bibnamefont
  {Braginski}},\ }\href@noop {} {\emph {\bibinfo {title} {The SQUID handbook:
  Applications of SQUIDs and SQUID systems}}}\ (\bibinfo  {publisher} {John
  Wiley \& Sons},\ \bibinfo {year} {2006})\BibitemShut {NoStop}%
\bibitem [{\citenamefont {Webb}(1977)}]{webb1977}%
  \BibitemOpen
  \bibfield  {author} {\bibinfo {author} {\bibfnamefont {R.~A.}\ \bibnamefont
  {Webb}},\ }\href {\doibase 10.1063/1.1134950} {\bibfield  {journal} {\bibinfo
   {journal} {Review of Scientific Instruments}\ }\textbf {\bibinfo {volume}
  {48}},\ \bibinfo {pages} {1585} (\bibinfo {year} {1977})}\BibitemShut
  {NoStop}%
\bibitem [{\citenamefont {Fan}\ \emph {et~al.}(1989)\citenamefont {Fan},
  \citenamefont {Heaney}, \citenamefont {Clarke}, \citenamefont {Newitt},
  \citenamefont {Wald}, \citenamefont {Hahn}, \citenamefont {Bielecki},\ and\
  \citenamefont {Pines}}]{fan1989}%
  \BibitemOpen
  \bibfield  {author} {\bibinfo {author} {\bibfnamefont {N.~Q.}\ \bibnamefont
  {Fan}}, \bibinfo {author} {\bibfnamefont {M.~B.}\ \bibnamefont {Heaney}},
  \bibinfo {author} {\bibfnamefont {J.}~\bibnamefont {Clarke}}, \bibinfo
  {author} {\bibfnamefont {D.}~\bibnamefont {Newitt}}, \bibinfo {author}
  {\bibfnamefont {L.~L.}\ \bibnamefont {Wald}}, \bibinfo {author}
  {\bibfnamefont {E.~L.}\ \bibnamefont {Hahn}}, \bibinfo {author}
  {\bibfnamefont {A.}~\bibnamefont {Bielecki}}, \ and\ \bibinfo {author}
  {\bibfnamefont {A.}~\bibnamefont {Pines}},\ }\href {\doibase
  10.1109/20.92510} {\bibfield  {journal} {\bibinfo  {journal} {IEEE
  Transactions on Magnetics}\ }\textbf {\bibinfo {volume} {25}},\ \bibinfo
  {pages} {1193} (\bibinfo {year} {1989})}\BibitemShut {NoStop}%
\bibitem [{\citenamefont {Greenberg}(1998)}]{greenberg1998}%
  \BibitemOpen
  \bibfield  {author} {\bibinfo {author} {\bibfnamefont {Y.~S.}\ \bibnamefont
  {Greenberg}},\ }\href {\doibase 10.1103/RevModPhys.70.175} {\bibfield
  {journal} {\bibinfo  {journal} {Reviews of Modern Physics}\ }\textbf
  {\bibinfo {volume} {70}},\ \bibinfo {pages} {175} (\bibinfo {year}
  {1998})}\BibitemShut {NoStop}%
\bibitem [{\citenamefont {Clarke}\ \emph {et~al.}(2007)\citenamefont {Clarke},
  \citenamefont {Hatridge},\ and\ \citenamefont {M{\"o}{\ss}le}}]{clarke2007}%
  \BibitemOpen
  \bibfield  {author} {\bibinfo {author} {\bibfnamefont {J.}~\bibnamefont
  {Clarke}}, \bibinfo {author} {\bibfnamefont {M.}~\bibnamefont {Hatridge}}, \
  and\ \bibinfo {author} {\bibfnamefont {M.}~\bibnamefont {M{\"o}{\ss}le}},\
  }\href {\doibase 10.1023/B:JOLT.0000029519.09286.c5} {\bibfield  {journal}
  {\bibinfo  {journal} {Annual Review of Biomedical Engineering}\ }\textbf
  {\bibinfo {volume} {9}},\ \bibinfo {pages} {389} (\bibinfo {year}
  {2007})}\BibitemShut {NoStop}%
\bibitem [{\citenamefont {Augustine}\ \emph {et~al.}(1998)\citenamefont
  {Augustine}, \citenamefont {TonThat},\ and\ \citenamefont
  {Clarke}}]{augustine1998}%
  \BibitemOpen
  \bibfield  {author} {\bibinfo {author} {\bibfnamefont {M.~P.}\ \bibnamefont
  {Augustine}}, \bibinfo {author} {\bibfnamefont {D.~M.}\ \bibnamefont
  {TonThat}}, \ and\ \bibinfo {author} {\bibfnamefont {J.}~\bibnamefont
  {Clarke}},\ }\href {\doibase 10.1016/S0926-2040(97)00103-3} {\bibfield
  {journal} {\bibinfo  {journal} {Solid State Nuclear Magnetic Resonance}\
  }\textbf {\bibinfo {volume} {11}},\ \bibinfo {pages} {139} (\bibinfo {year}
  {1998})}\BibitemShut {NoStop}%
\bibitem [{\citenamefont {McDermott}\ \emph {et~al.}(2004)\citenamefont
  {McDermott}, \citenamefont {Lee}, \citenamefont {Haken}, \citenamefont
  {Trabesinger}, \citenamefont {Pines},\ and\ \citenamefont
  {Clarke}}]{mcdermott2004}%
  \BibitemOpen
  \bibfield  {author} {\bibinfo {author} {\bibfnamefont {R.}~\bibnamefont
  {McDermott}}, \bibinfo {author} {\bibfnamefont {S.}~\bibnamefont {Lee}},
  \bibinfo {author} {\bibfnamefont {B.~t.}\ \bibnamefont {Haken}}, \bibinfo
  {author} {\bibfnamefont {A.~H.}\ \bibnamefont {Trabesinger}}, \bibinfo
  {author} {\bibfnamefont {A.}~\bibnamefont {Pines}}, \ and\ \bibinfo {author}
  {\bibfnamefont {J.}~\bibnamefont {Clarke}},\ }\href {\doibase
  10.1073/pnas.0402382101} {\bibfield  {journal} {\bibinfo  {journal} {PNAS}\
  }\textbf {\bibinfo {volume} {101}},\ \bibinfo {pages} {7857} (\bibinfo {year}
  {2004})}\BibitemShut {NoStop}%
\bibitem [{\citenamefont {Matlachov}\ \emph {et~al.}(2004)\citenamefont
  {Matlachov}, \citenamefont {Volegov}, \citenamefont {Espy}, \citenamefont
  {George},\ and\ \citenamefont {Kraus~Jr}}]{matlachov2004}%
  \BibitemOpen
  \bibfield  {author} {\bibinfo {author} {\bibfnamefont {A.~N.}\ \bibnamefont
  {Matlachov}}, \bibinfo {author} {\bibfnamefont {P.~L.}\ \bibnamefont
  {Volegov}}, \bibinfo {author} {\bibfnamefont {M.~A.}\ \bibnamefont {Espy}},
  \bibinfo {author} {\bibfnamefont {J.~S.}\ \bibnamefont {George}}, \ and\
  \bibinfo {author} {\bibfnamefont {R.~H.}\ \bibnamefont {Kraus~Jr}},\ }\href
  {\doibase 10.1016/j.jmr.2004.05.015} {\bibfield  {journal} {\bibinfo
  {journal} {Journal of Magnetic Resonance}\ }\textbf {\bibinfo {volume}
  {170}},\ \bibinfo {pages} {1} (\bibinfo {year} {2004})}\BibitemShut {NoStop}%
\bibitem [{\citenamefont {Buckenmaier}\ \emph {et~al.}(2017)\citenamefont
  {Buckenmaier}, \citenamefont {Rudolph}, \citenamefont {Back}, \citenamefont
  {Misztal}, \citenamefont {Bommerich}, \citenamefont {Fehling}, \citenamefont
  {Koelle}, \citenamefont {Kleiner}, \citenamefont {Mayer}, \citenamefont
  {Scheffler} \emph {et~al.}}]{buckenmaier2017}%
  \BibitemOpen
  \bibfield  {author} {\bibinfo {author} {\bibfnamefont {K.}~\bibnamefont
  {Buckenmaier}}, \bibinfo {author} {\bibfnamefont {M.}~\bibnamefont
  {Rudolph}}, \bibinfo {author} {\bibfnamefont {C.}~\bibnamefont {Back}},
  \bibinfo {author} {\bibfnamefont {T.}~\bibnamefont {Misztal}}, \bibinfo
  {author} {\bibfnamefont {U.}~\bibnamefont {Bommerich}}, \bibinfo {author}
  {\bibfnamefont {P.}~\bibnamefont {Fehling}}, \bibinfo {author} {\bibfnamefont
  {D.}~\bibnamefont {Koelle}}, \bibinfo {author} {\bibfnamefont
  {R.}~\bibnamefont {Kleiner}}, \bibinfo {author} {\bibfnamefont {H.~A.}\
  \bibnamefont {Mayer}}, \bibinfo {author} {\bibfnamefont {K.}~\bibnamefont
  {Scheffler}},  \emph {et~al.},\ }\href {\doibase 10.1038/s41598-017-13757-7}
  {\bibfield  {journal} {\bibinfo  {journal} {Scientific Reports}\ }\textbf
  {\bibinfo {volume} {7}},\ \bibinfo {pages} {13431} (\bibinfo {year}
  {2017})}\BibitemShut {NoStop}%
\bibitem [{\citenamefont {Ehnholm}\ \emph {et~al.}(1979)\citenamefont
  {Ehnholm}, \citenamefont {Ekstr{\"o}m}, \citenamefont {Loponen},\ and\
  \citenamefont {Soini}}]{ehnholm1979}%
  \BibitemOpen
  \bibfield  {author} {\bibinfo {author} {\bibfnamefont {G.~J.}\ \bibnamefont
  {Ehnholm}}, \bibinfo {author} {\bibfnamefont {J.~P.}\ \bibnamefont
  {Ekstr{\"o}m}}, \bibinfo {author} {\bibfnamefont {M.~T.}\ \bibnamefont
  {Loponen}}, \ and\ \bibinfo {author} {\bibfnamefont {J.~K.}\ \bibnamefont
  {Soini}},\ }\href {\doibase 10.1016/0011-2275(79)90071-7} {\bibfield
  {journal} {\bibinfo  {journal} {Cryogenics}\ }\textbf {\bibinfo {volume}
  {19}},\ \bibinfo {pages} {673} (\bibinfo {year} {1979})}\BibitemShut
  {NoStop}%
\bibitem [{\citenamefont {Pasquarelli}\ \emph {et~al.}(1996)\citenamefont
  {Pasquarelli}, \citenamefont {Del~Gratta}, \citenamefont {Della~Penna},
  \citenamefont {Di~Luzio}, \citenamefont {Pizzella},\ and\ \citenamefont
  {Romani}}]{pasquarelli1996}%
  \BibitemOpen
  \bibfield  {author} {\bibinfo {author} {\bibfnamefont {A.}~\bibnamefont
  {Pasquarelli}}, \bibinfo {author} {\bibfnamefont {C.}~\bibnamefont
  {Del~Gratta}}, \bibinfo {author} {\bibfnamefont {S.}~\bibnamefont
  {Della~Penna}}, \bibinfo {author} {\bibfnamefont {S.}~\bibnamefont
  {Di~Luzio}}, \bibinfo {author} {\bibfnamefont {V.}~\bibnamefont {Pizzella}},
  \ and\ \bibinfo {author} {\bibfnamefont {G.~L.}\ \bibnamefont {Romani}},\
  }\href {\doibase 10.1088/0031-9155/41/11/020} {\bibfield  {journal} {\bibinfo
   {journal} {Physics in Medicine and Biology}\ }\textbf {\bibinfo {volume}
  {41}},\ \bibinfo {pages} {2533} (\bibinfo {year} {1996})}\BibitemShut
  {NoStop}%
\bibitem [{\citenamefont {Pizzella}\ \emph {et~al.}(2001)\citenamefont
  {Pizzella}, \citenamefont {Della~Penna}, \citenamefont {Del~Gratta},\ and\
  \citenamefont {Romani}}]{pizzella2001}%
  \BibitemOpen
  \bibfield  {author} {\bibinfo {author} {\bibfnamefont {V.}~\bibnamefont
  {Pizzella}}, \bibinfo {author} {\bibfnamefont {S.}~\bibnamefont
  {Della~Penna}}, \bibinfo {author} {\bibfnamefont {C.}~\bibnamefont
  {Del~Gratta}}, \ and\ \bibinfo {author} {\bibfnamefont {G.~L.}\ \bibnamefont
  {Romani}},\ }\href {\doibase 10.1088/0953-2048/14/7/201} {\bibfield
  {journal} {\bibinfo  {journal} {Superconductor Science and Technology}\
  }\textbf {\bibinfo {volume} {14}},\ \bibinfo {pages} {R79} (\bibinfo {year}
  {2001})}\BibitemShut {NoStop}%
\bibitem [{\citenamefont {Poggio}\ \emph
  {et~al.}(2007{\natexlab{b}})\citenamefont {Poggio}, \citenamefont {Degen},
  \citenamefont {Rettner}, \citenamefont {Mamin},\ and\ \citenamefont
  {Rugar}}]{poggio2007b}%
  \BibitemOpen
  \bibfield  {author} {\bibinfo {author} {\bibfnamefont {M.}~\bibnamefont
  {Poggio}}, \bibinfo {author} {\bibfnamefont {C.~L.}\ \bibnamefont {Degen}},
  \bibinfo {author} {\bibfnamefont {C.~T.}\ \bibnamefont {Rettner}}, \bibinfo
  {author} {\bibfnamefont {H.~J.}\ \bibnamefont {Mamin}}, \ and\ \bibinfo
  {author} {\bibfnamefont {D.}~\bibnamefont {Rugar}},\ }\href {\doibase
  10.1063/1.2752536} {\bibfield  {journal} {\bibinfo  {journal} {Applied
  Physics Letters}\ }\textbf {\bibinfo {volume} {90}},\ \bibinfo {pages}
  {263111} (\bibinfo {year} {2007}{\natexlab{b}})}\BibitemShut {NoStop}%
\bibitem [{\citenamefont {Nichol}\ \emph {et~al.}(2012)\citenamefont {Nichol},
  \citenamefont {Hemesath}, \citenamefont {Lauhon},\ and\ \citenamefont
  {Budakian}}]{nichol2012}%
  \BibitemOpen
  \bibfield  {author} {\bibinfo {author} {\bibfnamefont {J.~M.}\ \bibnamefont
  {Nichol}}, \bibinfo {author} {\bibfnamefont {E.~R.}\ \bibnamefont
  {Hemesath}}, \bibinfo {author} {\bibfnamefont {L.~J.}\ \bibnamefont
  {Lauhon}}, \ and\ \bibinfo {author} {\bibfnamefont {R.}~\bibnamefont
  {Budakian}},\ }\href {\doibase 10.1103/PhysRevB.85.054414} {\bibfield
  {journal} {\bibinfo  {journal} {Physical Review B}\ }\textbf {\bibinfo
  {volume} {85}},\ \bibinfo {pages} {054414} (\bibinfo {year}
  {2012})}\BibitemShut {NoStop}%
\bibitem [{\citenamefont {{Quantum Design
  Inc.}}(2018)}]{website:QuantumDesign}%
  \BibitemOpen
  \bibfield  {author} {\bibinfo {author} {\bibnamefont {{Quantum Design
  Inc.}}},\ }\href
  {https://www.qdusa.com/sitedocs/appNotes/squids/1052-202.pdf} {\enquote
  {\bibinfo {title} {Squid application note 1052-202a: Coupling magnetic
  signals to a squid amplifier},}\ } (\bibinfo {year} {2018})\BibitemShut
  {NoStop}%
\bibitem [{\citenamefont {Garwood}\ and\ \citenamefont
  {DelaBarre}(2001)}]{garwood2001}%
  \BibitemOpen
  \bibfield  {author} {\bibinfo {author} {\bibfnamefont {M.}~\bibnamefont
  {Garwood}}\ and\ \bibinfo {author} {\bibfnamefont {L.}~\bibnamefont
  {DelaBarre}},\ }\href {\doibase 10.1006/jmre.2001.2340} {\bibfield  {journal}
  {\bibinfo  {journal} {Journal of Magnetic Resonance}\ }\textbf {\bibinfo
  {volume} {153}},\ \bibinfo {pages} {155} (\bibinfo {year}
  {2001})}\BibitemShut {NoStop}%
\bibitem [{\citenamefont {O'Dell}(2013)}]{odell2013}%
  \BibitemOpen
  \bibfield  {author} {\bibinfo {author} {\bibfnamefont {L.~A.}\ \bibnamefont
  {O'Dell}},\ }\href {\doibase 10.1016/j.ssnmr.2013.10.003} {\bibfield
  {journal} {\bibinfo  {journal} {Solid State Nuclear Magnetic Resonance}\
  }\textbf {\bibinfo {volume} {55}},\ \bibinfo {pages} {28} (\bibinfo {year}
  {2013})}\BibitemShut {NoStop}%
\bibitem [{\citenamefont {Claridge}(2016)}]{claridge2016}%
  \BibitemOpen
  \bibfield  {author} {\bibinfo {author} {\bibfnamefont {T.~D.~W.}\
  \bibnamefont {Claridge}},\ }\href@noop {} {\emph {\bibinfo {title}
  {High-resolution NMR techniques in organic chemistry}}},\ Vol.~\bibinfo
  {volume} {27}\ (\bibinfo  {publisher} {Elsevier},\ \bibinfo {year}
  {2016})\BibitemShut {NoStop}%
\bibitem [{\citenamefont {Choi}\ \emph {et~al.}(2016)\citenamefont {Choi},
  \citenamefont {Kim}, \citenamefont {Jeong}, \citenamefont {Kim},
  \citenamefont {Park},\ and\ \citenamefont {Park}}]{choi2016}%
  \BibitemOpen
  \bibfield  {author} {\bibinfo {author} {\bibfnamefont {W.~S.}\ \bibnamefont
  {Choi}}, \bibinfo {author} {\bibfnamefont {M.~J.}\ \bibnamefont {Kim}},
  \bibinfo {author} {\bibfnamefont {I.~W.}\ \bibnamefont {Jeong}}, \bibinfo
  {author} {\bibfnamefont {D.~E.}\ \bibnamefont {Kim}}, \bibinfo {author}
  {\bibfnamefont {H.~C.}\ \bibnamefont {Park}}, \ and\ \bibinfo {author}
  {\bibfnamefont {K.~H.}\ \bibnamefont {Park}},\ }\href {\doibase
  10.1016/j.nima.2016.03.078} {\bibfield  {journal} {\bibinfo  {journal}
  {Nuclear Instruments and Methods in Physics Research Section A: Accelerators,
  Spectrometers, Detectors and Associated Equipment}\ }\textbf {\bibinfo
  {volume} {822}},\ \bibinfo {pages} {15} (\bibinfo {year} {2016})}\BibitemShut
  {NoStop}%
\bibitem [{\citenamefont {van Waarde}\ \emph {et~al.}(2016)\citenamefont {van
  Waarde}, \citenamefont {Benningshof},\ and\ \citenamefont
  {Oosterkamp}}]{waarde2016}%
  \BibitemOpen
  \bibfield  {author} {\bibinfo {author} {\bibfnamefont {B.}~\bibnamefont {van
  Waarde}}, \bibinfo {author} {\bibfnamefont {O.~W.~B.}\ \bibnamefont
  {Benningshof}}, \ and\ \bibinfo {author} {\bibfnamefont {T.~H.}\ \bibnamefont
  {Oosterkamp}},\ }\href {\doibase 10.1016/j.cryogenics.2016.06.014} {\bibfield
   {journal} {\bibinfo  {journal} {Cryogenics}\ }\textbf {\bibinfo {volume}
  {78}},\ \bibinfo {pages} {74} (\bibinfo {year} {2016})}\BibitemShut {NoStop}%
\end{thebibliography}%
\bibliographystyle{apsrev4-1}

\end{document}